# *In silico* study on the contribution of the follicular route to dermal permeability of small molecules.

Daniel Sebastia-Saez[1], Guoping Lian[2], Tao Chen[1]

[1]School of Chemistry and Chemical Engineering, University of Surrey, Guildford, United Kingdom

[2]Unilever R&D Colworth, Bedford, United Kingdom

*Corresponding author: Daniel Sebastia-Saez (j.sebastiasaez@surrey.ac.uk)

## Abstract

**Purpose**. This study investigates *in silico* the contribution of the hair follicle to the overall dermal permeability of small molecules, as published experimental work provides inconclusive information on whether the follicular route favours the permeation of hydrophobic or hydrophilic permeants.

**Method**. A study is conducted varying physico-chemical parameters of permeants such as lipophilicity, molecular weight and protein binding. The simulated data is compared to published experimental data to discuss how those properties can modulate the contribution of the hair follicle to the overall dermal permeation.

**Results**. The results indicate that the contribution of the follicular route to dermal permeation can range from negligible to notable depending on the combination of lipophilic/hydrophilic properties of the substance filling the follicular route and the permeant.

**Conclusion**. Characterisation of the substance filling the follicular route is required for analysing the experimental data of dermal permeation of small molecules, as changes between *in vivo* and *in vitro* due to handling of samples and cessation of vital functions can modify the contribution of the follicular route to overall dermal permeation, hence hindering data interpretation.

## Abbreviations

### Latin characters

| | |
|---|---|
| $C, C_o$ | Concentration [mol·m$^{-3}$]. The subscript $o$ denotes the infinite dose concentration |
| $CFR$ | Contribution of the follicular route [%] |
| $D_{o/w}$ | Distribution coefficient. |
| $D, D_{SC}, D_{VE}, D_{De}, D_{se}, D_w$ | Diffusion coefficient [m$^2$·s$^{-1}$]: Generic (no subscript), in the stratum corneum ($SC$), viable epidermis ($VE$), dermis ($De$), sebum ($se$) and water ($w$). |
| $f_u, f_{non}$ | Fraction unbound ($u$) to protein and non-ionised ($non$) fraction |
| $\vec{J}$ | Fick's diffusive flux [mol·m$^{-2}$·s$^{-1}$] |
| $k_b$ | Boltzmann constant: 1.3806×10$^{-23}$ [J·K$^{-1}$] |
| $K_p$ | Permeability of the stratum corneum [m·s$^{-1}$] |
| $K_{o/w}, K_{SC/w}, K_{VE/w}, K_{De/w}, K_{se/w}, K_{pro/w}, K_{lip/w}$ | Partition coefficients: Octanol to water ($o/w$), stratum corneum to water ($SC/w$), viable epidermis to water ($VE/w$), dermis to water ($De/w$) and sebum to water ($se/w$), |
| | protein to water ($pro/w$) and lipid to water ($lip/w$). |
| $MW$ | Molecular weight [Da] |
| $P_{open}, P_{blocked}, P_{impaired}$ | Permeability [m·s$^{-1}$] upon open, blocked and impaired follicular route. |
| $r$ | Molecular radius [Å] |
| $R^2$ | Coefficient of determination |
| $t$ | Time [s] |
| $T$ | Temperature [K] |
| **Greek characters** | |
| $\delta$ | Thickness of the stratum corneum [μm] |
| $\eta$ | Dynamic viscosity [Pa·s] |
| $\rho_{pro}, \rho_{lip}, \rho_w$ | Density of protein ($pro$), lipid ($lip$) and water ($w$). |
| $\varphi_{pro}, \varphi_{lip}, \varphi_w$ | Volume fraction of protein ($pro$), lipid ($lip$) and water ($w$). |

## Introduction

Understanding the mechanisms behind dermal permeation of chemicals is of high importance for the pharmaceutical and cosmetic industries, as well as for risk assessment of chemical exposure. Reported experimental studies have shown that the follicular route (FR) contributes notably to dermal permeation (Knorr et al., 2006). This can be explained as the FR offers large surface area for mass transport due to an enfolding around the hair that extends deeply into the dermis (De), hence by-passing the stratum corneum (SC) as the skin's main barrier. Published experimental data is inconclusive on the mechanisms behind the role of the FR on the overall dermal permeability. Specifically, it is inconclusive on whether the FR favours the permeation of hydrophilic or lipophilic compounds, with implications on the general understanding of the delivery and safety of dermal formulations. Table 1 summarises studies on dermal permeability through the FR published in the last two decades. While these studies showed that the FR plays an important role, further clarification is required regarding the effect of the physico-chemical properties of the permeants, such as lipophilicity and molecular weight. For example, Otberg et al. (2008) found that the FR increases the concentration of caffeine (a hydrophilic compound) in plasma in their *in vivo* experiments, suggesting that the FR enhances the permeation of hydrophilic compounds *in vivo*. However, it is well known that continuous secretion of sebum by the sebaceous gland happens *in vivo*, and therefore the FR should not have such a noticeable contribution to the permeability of caffeine. Teichmann et al. (2007) reported *in vivo* experiments and observed that a lipophilic compound (curcumin, $\log K_{o/w}$=3.30) permeated into the skin through the FR and the intercellular lipidic phase in the SC but not through the hydrophilic keratinocytes. Their experimental observations are thus in accordance with the postulated lipophilic character of the FR *in vivo*.

Similarly, inconclusive results can be observed in the literature when comparing experiments carried out *in vitro* and *ex vivo*, where the sebaceous gland has ceased to produce sebum. Specimens are often soaked in aqueous solutions (e.g., PBS) prior to *ex vivo* and *in vitro* permeation experiments, which together with freezing and thawing of samples can contribute to altering the substance filling the FR. These alterations could lead to the permeation of hydrophilic compounds being favoured *in vitro* or *ex vivo* as opposed to *in vivo*. For example, Genina et al. (2002) studied the permeation of a lipophilic compound (Dye Indigo Carmine, $\log K_{o/w}$ = 3.72) and a hydrophilic compound (Dye Indocyanina Green, $\log K_{o/w}$ = -0.79) and showed that the hydrophilic compound permeated faster than the lipophilic compound *in vitro*. This would be expected according to the postulated changes taking place during *in vitro* sample preparation. Grams and Bouwstra (2002) however, detected the presence of lipophilic compounds in the FR after their *ex vivo* permeation experiments, but not hydrophilic ones, suggesting that the FR favours the permeation of lipophilic compounds even when sebum secretion has ceased. These results could be explained as arguably *ex vivo* experiments mimic *in vivo* conditions more closely than *in vitro* tests (i.e., sebum still present in the FR).

Mohd et al. (2016) and Frum et al. (2007) directly compared the contribution of the FR on the overall skin permeation for a set of compounds with a sufficient range of $\log K_{o/w}$ *in vitro*. However, the contribution of the FR was obtained using different methodologies in both studies. While Frum et al. (2007) used the skin sandwich technique in their experiments, Mohd et al. (2016) opted for wax plugging half of the hair follicles in their skin samples. Mohd et al. (2016) observed that the contribution of the FR to dermal permeability was noticeably greater for hydrophilic compounds. Their results suggest thus that the FR favours the permeation of hydrophilic compounds *in vitro*. Frum et al. (2007) also provided direct comparison between the contribution of the FR against $\log K_{o/w}$ for a set of chemicals *in vitro*. In their study, the opposite trend was observed at low values of $\log K_{o/w}$, with a turning point at $\log K_{o/w}$=1.25. Frum et al. (2007) conclude that not only the lipophilicity, but other phenomena such as hydrogen bonding play an important role on the contribution of the FR on dermal permeability.

Published experimental data suggested that both the molecular weight and the lipophilicity play a substantial role. However, further investigation on how these and other parameters are involved in the process is needed. Here, we use a recently reported 2-D mechanistic *in silico* skin permeation mathematical model. The model has been validated recently in the literature for a variety of cases (Kattou et al., 2017; Sebastia-Saez et al., 2023). This methodology allows carrying out a study for a set of chemical compounds and examining the effects of the parameters involved.

*Table 1 Summary of published experimental studies on the effect of the FR on transdermal permeation of small molecules. [a]The contribution of the FR on overall absorption was not reported in the experiment; rather, it was determined to be very small by an in silico study (Kattou et al., 2017) when the model provided good agreement with the experimental plasma measurement. [b]Two values were obtained depending on the pH of the buffer solution used.*

| Reference | Test Condition | Compound(s) | MW [Da] | $\log K_{o/w}$ | Findings |
|---|---|---|---|---|---|
| Otberg et al. (2008) | *In vivo* | Caffeine | 194 | -0.07 | The FR contributes to increased concentration in blood plasma[a]. |
| Teichmann et al. (2005) | *In vivo* | Sodium Fluorescein (SF) | 376 | -0.61 | SF was recovered from follicular casts after experiment (3% of the dose was in the FR, 51% in corneocytes and the remainder was absorbed). |
| Teichmann et al. (2007) | *In vivo* | Curcumin | 368 | 3.30 | Curcumin permeates into the skin through lipid and FR. |
| Schwartz et al. (2011) | *In vivo* | Zinc pyrithione (ZP) | 317 | 0.90 | Notable amount of ZP detected in FR post-assay |
| Ossadnik et al. (2007) | *In vivo* | Brilliant green solution | 484 | 2.02 | Brilliant green solution detected in FR post-assay. |
| Grams and Bouwstra (2002) | *Ex vivo* | Bodipy FLc5 | 320 | 1.20 | Bodipy FLc5 and Bodipy 564/570C5 detected in FR post-assay, but not Oregon Green 488 |
| | | Bodipy 564/570C5 | 463 | 3.00 | |
| | | Oregon Green 488 | 509 | -2.50 | |
| Chandrasekaran et al. (2016) | *In vitro* | Magnesium | 23 | -1.10 | Magnesium permeates better into the skin when the FR is not blocked. |
| Genina et al. (2002) | *In vivo & in vitro* | Dye Indigo Carmine (IG) | 466 | 3.72 | Both substances were present in the FR after *in vivo* assays, but ICG permeates faster *in vitro*. |
| | | Dye Indocyanine Green (ICG) | 775 | -0.29 | |
| Frum et al. (2007) | *In vitro* | Estradiol | 272 | 2.29 | The FR does not show a clear relationship with the lipophilicity of the compound in this study. |
| | | Corticosterone | 346 | 1.94 | |
| | | Hydrocortisone | 362 | 1.60 | |
| | | Aldosterone | 360 | 1.08 | |
| | | Cimetidine | 252 | 0.40 | |
| | | Deoxyadenosine | 251 | -0.55 | |
| | | Adenosine | 267 | -1.05 | |
| Ogiso et al. (2002) | *In vitro* | Ketoprofen (KP) | 254 | 3.12 | Flux of KP much higher than that observed for MT and 5FU. Histological observations confirmed permeation had occurred through the FR. |
| | | Melatonin (MT) | 232 | 1.18 | |
| | | Fluorouracil (5FU) | 130 | -0.89 | |
| Mohd et al. (2016) | *In vitro* | Fluorescein isothiocyanate-dextran | 4000 | -0.77 | The follicular pathway clearly favoured the permeation of hydrophilic compounds, with a direct relationship between log $K_{o/w}$ and contribution of the FR. |
| | | Calcein sodium salt | 644 | -3.50 | |
| | | Fluorescein sodium salt | 376 | -0.61 | |
| | | Isosorbide dinitrate | 236 | 1.23 | |
| | | Lidocaine hydrochloride | 234 | -0.9/1.4[b] | |
| | | Aminopyrine | 231 | -1/0.98[b] | |
| | | Ibuprofen | 206 | 1.9/1.3[b] | |
| | | Butyl paraben | 194 | 3.50 | |
| | | Isosorbide mononitrate | 191 | -0.15 | |
| Essa et al. (2002) | *In vitro* | Estradiol | 272 | 2.29 | Greater contribution of the FR for hydrophilic than lipophilic compounds. |
| | | Mannitol | 182 | -2.47 | |
| | | Liposomes | - | | |

## Materials and Methods

The mechanistic mathematical model used in this work consists in a 2-D geometric representation of the skin layers, complemented with state-of-the-art QSPRs to calculate the diffusion and partition coefficients. The model is an adaptation of a previously published model (Kattou et al., 2017; Sebastia-Saez et al., 2023), to accommodate a configuration with an infinite dose in the vehicle and a sink in the receiver, used frequently in skin permeability lab work. Dimensions and boundary layers are represented in Figure 1 for an open FR. Cases with open, impaired, and closed FR were run in this study. In impaired FR cases, flux through the concentration boundary of the FR was restricted to half that of the open FR cases to mimic the experimental conditions of Mohd et al. (2016), who blocked half of the hair follicles in their experimental samples. In closed FR cases, the concentration boundary of the FR was replaced by a no flux boundary to suppress the flow from the vehicle directly into the FR.

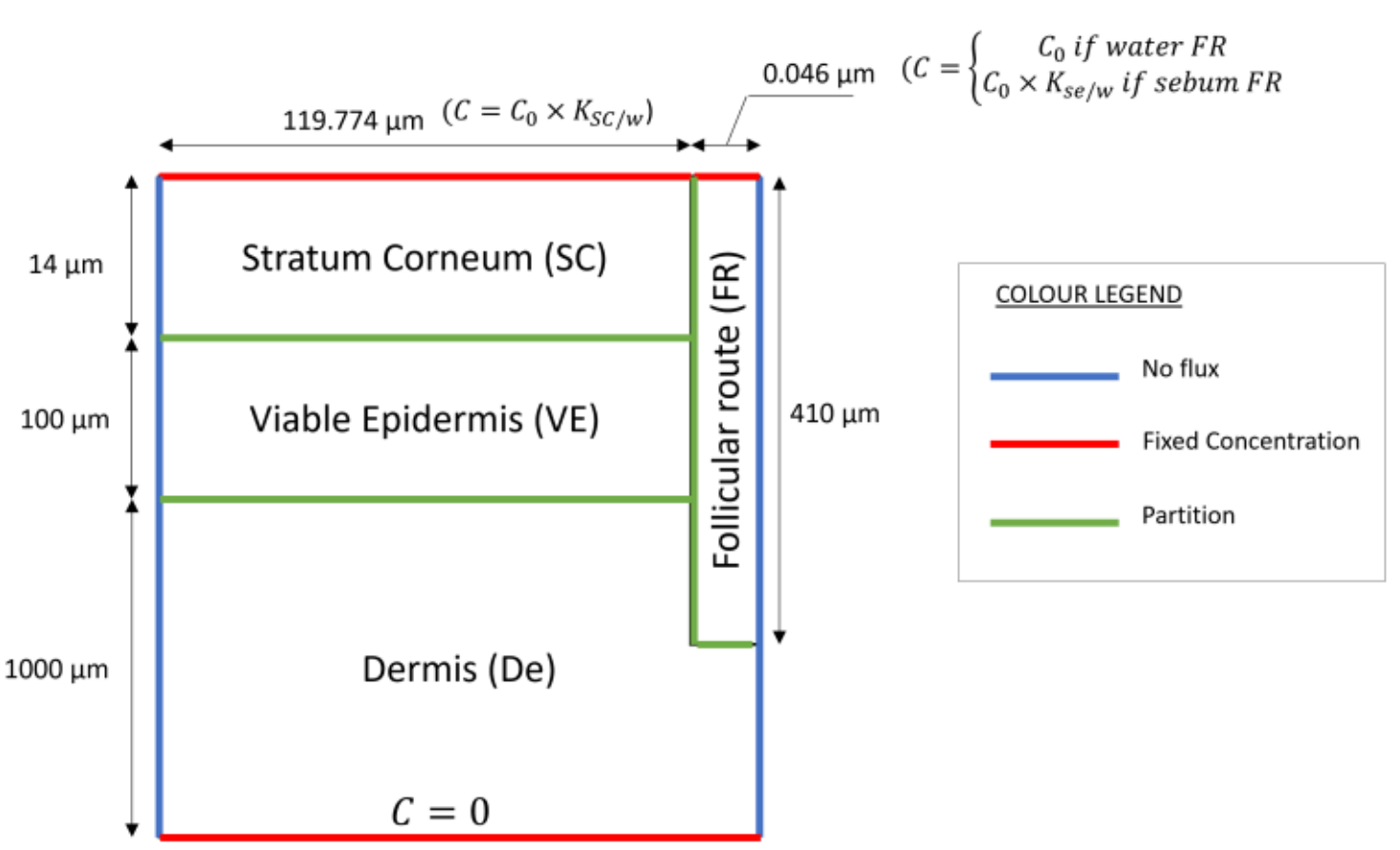

*Figure 1 Schematic illustration of the computational domain including colour-coded boundary layers and dimensions. The figure includes only the boundaries for the case of open FR. Fixed concentration boundaries are included for the cases with a FR filled with water or sebum. Not to scale. Dimensions obtained from Kattou et al. (2017).*

A fixed concentration $C_0 = 1$ mol/m$^3$ was fixed for the aqueous vehicle with infinite dose, as the concentration does not affect the permeability of the setup. The model was implemented using the commercial software COMSOL Multiphysics v6.1 and solves the diffusion equation with no convection.

$$\frac{\partial C}{\partial t} + \nabla \cdot \vec{J} = 0 \tag{1}$$

where $c$ denotes concentration, $t$ denotes time, and $\vec{J}$ is Fick's diffusive flux: $\vec{J} = -D\nabla C$, with $D$ being the diffusion coefficient.

Diffusion and partition coefficients are needed for each layer to run the model. Recently reported QSPR models with improved coefficients of determination were used when available to obtain the partition coefficients. These are summarised in Table 2. The present model includes a homogeneous SC (Wang et al., 2010). The literature suggests that De and VE have similar multiphase compositions (Dancik et al., 2013; Kretsos et al., 2008), and therefore the diffusion and partition properties are assumed to be similar (Kattou et al., 2017).

With the purpose of checking the validity of the homogeneous SC assumption, the simulation setup published by Kattou et al. (2017), different from the setup described above, was implemented in COMSOL Multiphysics v6.1 to replicate their results for dermal absorption of a finite-dose of caffeine with systemic circulation and clearance. Their setup and equations were implemented with the only difference of the homogenised SC, while Kattou et al. (2017) used the bricks and mortar layout.

*Table 2 Summary of expressions used to calculate partition parameters. [a]Volume fractions and densities are respectively $\varphi_{pro} = 0.1476$, $\varphi_{lip} = 0.0671$ $\varphi_w = 0.7853$, $\rho_{pro} = 1.37$ g/cm$^3$, $\rho_{lip} = 0.90$ g/cm$^3$, $\rho_w = 1.00$ g/cm$^3$ and $\rho_{SC} = 1.05$ g/cm$^3$ (Wang et al., 2010). [b]Partition coefficients calculated as $K_{pro/w} = 4.2 \times K_{o/w}^{0.31}$ and $K_{lip/w} = K_{o/w}^{0.69}$ (Wang et al., 2010).*

| **Parameter** | **Expression** |
| --- | --- |
| Partition coefficient SC to water[a,b] ($R^2 = 0.89$) (Wang et al., 2010) | $K_{SC/w} = \left(\varphi_{pro} \times \frac{\rho_{pro}}{\rho_w} \times K_{pro/w} + \varphi_{lip} \times \frac{\rho_{lip}}{\rho_w} \times K_{lip/w} + \varphi_w\right) \times \frac{\rho_w}{\rho_{SC}}$ |
| Partition coefficient VE to water and De to water (Kretsos et al., 2018; Ibrahim et al., 2012) | $K_{VE/w} = K_{De/w} = 0.7 \times \left(0.68 + \frac{0.32}{f_u} + 0.025 \times f_{non} \times K_{o/w}^{0.7}\right)$ |

And the expressions used to calculate the diffusion coefficients are gathered in Table 3.

*Table 3 Summary of expressions used to calculate diffusion parameters in SI units. [a]The following QSPR (Mitragotri, 2002) was used to calculate the permeability of the SC: $K_p = 5.6 \times 10^{-6} K_{o/w}^{0.7} e^{-0.46r^2}$. Thickness of the SC is $\delta = 14$ μm as indicated in Figure 1 (Kattou et al., 2017). Partition coefficient SC to water $K_{SC/w}$ calculated as indicated in Table 2. [b]Partition coefficient lipid to water was $K_{mw} = \frac{\rho_{lip}}{\rho_w} K_{o/w}^{0.69}$(Chen et al., 2015). [c]Molecular radius $r = \sqrt[3]{3/4\pi \times 0.9087 \times MW}$ (Mitragotri et al., 1999). [d]Boltzmann constant $k_b = 1.380649 \times 10^{-23}$ J/K. Temperature $T = 20$°C. Dynamic viscosity of water $\eta = 1.0016$ mPa·s.*

| **Parameter** | **Expression** |
|---|---|
| Diffusion coefficient in SC[a] | $D_{SC} = \frac{K_p \times \delta}{K_{SC/w}}$ |
| Diffusion coefficient in VE and De[b] (Kattou et al., 2017) | $D_{VE} = \frac{10^{-8.15-0.655\times\log MW}}{0.68 + \frac{0.32}{f_u} + 0.025 \times f_{non} \times K_{mw}}$ |
| Diffusion coefficient in sebum ($R^2 = 0.92$)[c] (Yang et al., 2019) | $D_{se} = 2.48 \times 10^{-4} e^{-0.42\times r^2}$ |
| Diffusion coefficient in water (Stokes equation) (Einstein, 1905) | $D_w = \frac{k_b \times T}{6 \times \pi \times \eta \times r}$ |

The contribution of the follicular route (CFR) to skin permeability was calculated as:

$$CFR[\%] = \frac{P_{open} - P_{impaired}}{P_{open}} \times 100 \qquad (2)$$

where $P$ represents permeability and is obtained by dividing the flux at steady state $J$ over the concentration in the donor $C_0$ as $P = J/_{C_0}$. The flux was obtained from the simulations when steady state was reached (variations below 0.1% across the domain). The calculation of one value of the parameter $CFR$ requires two simulations: One with the FR open and another with the FR impaired. The simulations will be carried out for the cases of a hydrophilic (water) and lipophilic (sebum) FR, mimicking respectively, an *in vitro* and an *in vivo* scenario.

For a water FR, simulations will be carried out for 11 values of $\log K_{o/w}$ in the range [-3.70, 5.49] and two values of molecular weight (i.e., $MW = 50$ Da and 285 Da). These cover the range of values of a large number of commercial cosmetic and therapeutic drugs (Chen et al., 2013; Yang et al., 2018). The values of unbound $f_u$ and non-ionized $f_{non}$ fraction were tested at ~0 and 1. For each value of $\log K_{o/w}$, all possible combinations of molecular weight $MW$, unbound fraction $f_u$ and non-ionised fraction $f_{non}$ were simulated. This made a total of 88 virtual compounds (176 simulations for both blocked and unblocked FR). This methodology allowed comparison between the trends obtained by Mohd et al. (2016) and Frum et al. (2007) and the simulations in this study.

Further, a check was conducted on the uncertainty introduced by QSPRs used to compute the permeability of the SC due to low coefficient of determination ($R^2 = 0.58$) (Mitragotri, 2002). The contribution of the FR against $\log K_{o/w}$ was obtained for a set of chemicals with reported experimental measurements of the permeability (Chen et al., 2013) and compared with the QSPR prediction. The parameters of the chemicals implemented in these simulations are summarised in Table 4. An additional set of 40 simulations (with water FR) was carried out to perform this check. Values of $f_u = 1$ and $f_{non} = 1$ were assumed in this set of simulations to obtain the diffusion and partition coefficients for De and VE.

*Table 4 Parameters of chemical compounds used in uncertainty check.*

| **Compound** | $\log K_{o/w}$ | $MW$ **[Da]** | **Experimental** $\log K_p$ **[cm/s] [Chen et al., 2013]** |
|---|---|---|---|
| Sucrose | -3.70 | 342 | -8.84 |
| Aspartic acid | -3.47 | 133 | -7.43 |
| Lysine | -3.05 | 146 | -6.87 |
| Histidine | -2.90 | 155 | -7.82 |
| Urea | -2.11 | 61 | -7.39 |
| Methanol | -0.77 | 32 | -6.38 |
| Ethanol | -0.31 | 46 | -7.06 |
| Nicotinic acid | 0.36 | 123 | -8.18 |
| Barbital | 0.65 | 184 | -7.51 |
| Aldosterone | 1.08 | 360 | -7.86 |
| Codeine | 1.19 | 299 | -7.09 |
| Pentanol | 1.51 | 88 | -5.78 |
| Hexanol | 2.03 | 102 | -5.44 |
| Salicylic acid | 2.26 | 138 | -5.07 |
| Hydrocortisone methylsuccinate | 2.60 | 477 | -7.23 |
| Naproxen | 3.18 | 230 | -4.97 |
| Testosterone | 3.32 | 288 | -6.21 |
| Progesterone | 3.87 | 314 | -5.43 |
| Diclofenac | 4.51 | 296 | -5.30 |
| Hydrocortisone octanoate | 5.49 | 489 | -4.76 |

Cases with sebum FR were run for 9 chemicals with $\log K_{se/w}$ measured experimentally by Yang et al. (2018) (Table 5). The set includes the chemicals where $\log K_{o/w} = \log D_{o/w}$ to avoid the effect of ionization (e.g., a negative $\log K_{se/w}$ was measured for procaine, a lipophilic compound, due to ionization). A set of 18 simulations was performed for sebum FR. The compounds were assumed to be fully unbound (i.e., $f_u = 1$) and fully non-ionised (i.e., $f_{non} = 1$) in this set of simulations.

*Table 5. Parameters of the compounds tested for sebum FR (Yang et al., 2018).*

| **Compound** | $\log D_{o/w} = \log K_{o/w}$ | **Experimental** $\log K_{se/w}$ **[Yang et al., 2018]** | $MW$ **[Da]** | $pH$ |
|---|---:|---:|---:|---:|
| Theophylline | -0.54 | -1.00 | 180 | 5.5 |
| Theobromine | -0.36 | -1.55 | 180 | 5.5 |
| Caffeine | -0.15 | -0.77 | 194 | 5.5 |
| Vanillin | 1.23 | 0.83 | 152 | 5.5 |
| Thiabendazole | 2.48 | 1.77 | 201 | 5.5 |
| Testosterone | 3.27 | 2.26 | 288 | 5.5 |
| Thymol | 3.40 | 2.62 | 150 | 5.2 |
| Benzophenone | 3.43 | 2.61 | 182 | 3.8 |
| Octopirox | 3.50 | 3.01 | 238 | 5.5 |

## Results and Discussion

The simulations run to check the validity of the homogeneous SC assumption are gathered in Figure 2. The experimental results of Otberg et al. (2008) used by Kattou et al. (2017) for validation are also included in the graph. The simulations with homogenized SC used the experimental SC permeability of Johnson et al. (1997) to calculate the diffusion coefficient in the SC ($D_{SC} = 4.12 \times 10^{-15}$ m$^2$/s), Abraham and Martins (2004) ($D_{SC} = 4.73 \times 10^{-15}$ m$^2$/s) and the QSPR prediction by Mitragotri (2002) ($D_{SC} = 4.73 \times 10^{-15}$ m$^2$/s). The three simulated data series compare well with the simulated results using the bricks and mortar approach (Kattou et al., 2017), hence providing validation of the homogeneous SC assumption. Errors in the comparison between all the simulated results and the experiments of Otberg et al. (2018) are attributed to inaccuracies in the values presented in Kattou et al. (2017) for the parameters of the ODE governing the concentration of permeant in blood.

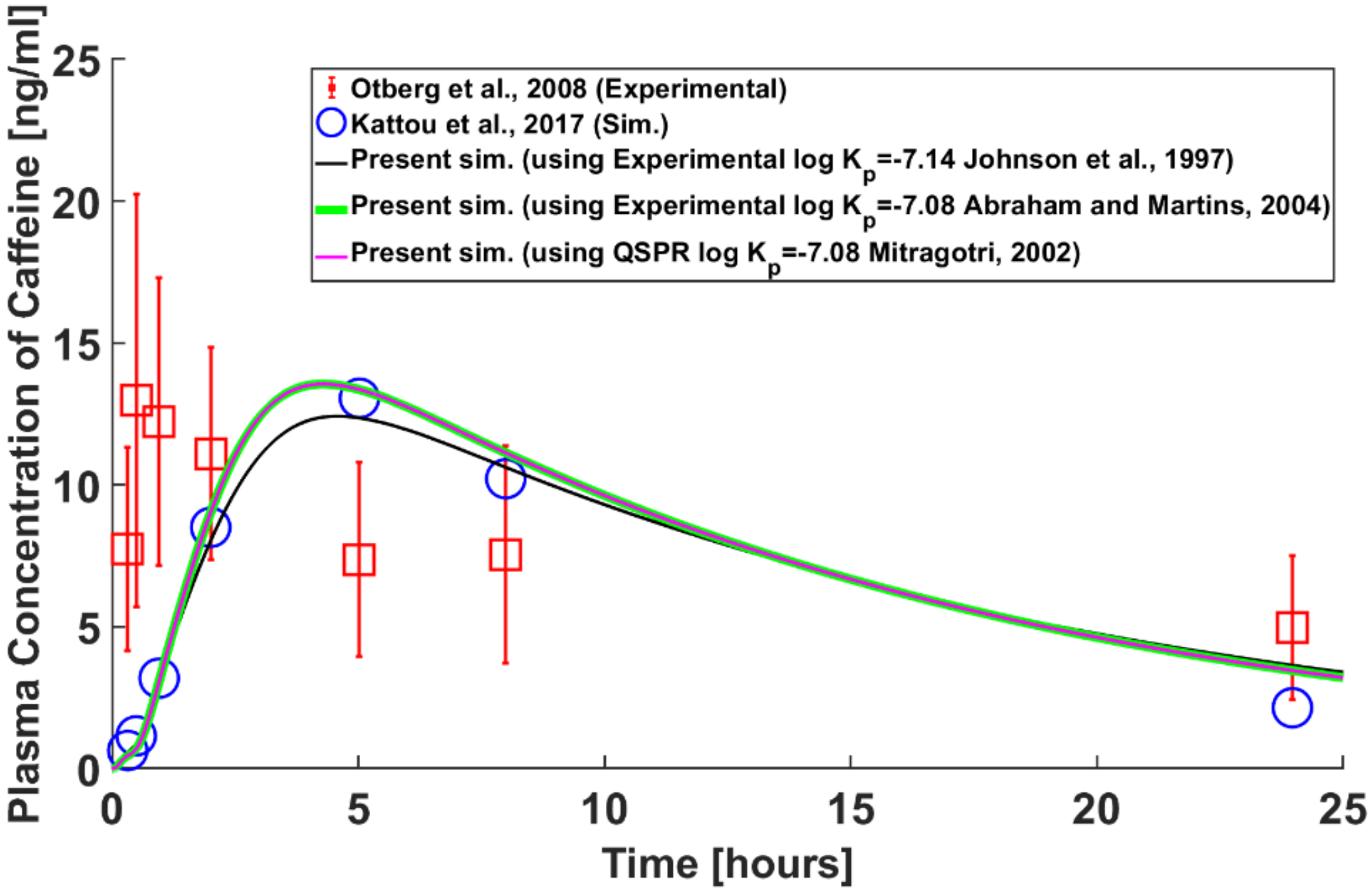

*Figure 2 Comparison between results obtained using the 'bricks and mortar' SC layout (Kattou et al., 2017) for dermal absorption of caffeine with blood clearance and the replicated results using the homogenized SC assumption (Present sim).*

The results for a water FR are in Figure 3. $CFR$ is plotted against $\log K_{o/w}$ to facilitate comparison against reported experimental data. An inverse relationship between the contribution of the FR and the octanol-to-water partition coefficient $\log K_{o/w}$ can be observed, confirming that a water FR favours the permeation of hydrophilic virtual compounds. Three areas can be distinguished on the graph. For hydrophilic compounds (i.e., $\log K_{o/w} < 1$), the contribution of the follicular path to the permeability of the skin is close to 50% regardless of molecular weight, percentage of protein binding and percentage of ionisation. Thus, most of the applied dose would permeate into the sink via the FR, hence by-passing the lipidic route in the SC. This is in accordance with the data of Mohd et al. (2016), who observed values of the $CFR$ of around 50% for highly hydrophilic compounds in their experiments wax-plugging half of the hair follicles in their samples, hence suggesting a notable contribution of the FR. For strongly lipophilic molecules (i.e., $\log K_{o/w} > 6$), the effect is the opposite; the proportion of the dose crossing into the sink via the FR is negligible. For molecules with $\log K_{o/w}$ ranging from 1 through 6, a balance can be found between the epidermal and the FR.

The data shows that the molecular weight has an important impact, although the trend remains unchanged for both values of $MW$ tested. For a given value of the compound's lipophilicity, the contribution of the FR increases with the molecular weight regardless of protein binding and ionisation. Experimental observations in the literature suggest an inverse relationship between the permeability of the SC and molecular weight of the compounds (Mitragotri, 2002), which could result in the FR contributing more than the epidermal route to overall dermal permeation for increased $MW$. This is consistent with this simulated data, as the contribution of the FR is smaller for 50 Da than for 285 Da virtual compounds. No effect of unbound $f_u$ and non-ionized fraction $f_{non}$ was observed. Although the latter two parameters have an important weight on the calculation of the diffusion and partition properties of the viable epidermis and dermis, the resistance to diffusion posed by the stratum corneum results in protein binding and ionisation having a negligible effect on the overall permeation rate.

The simulated data is thus in accordance with the hypothesis by which physico-chemical changes might occur *in vitro* that result in the FR favouring the permeation of hydrophilic compounds. The trend agrees with the experimental observations of Mohd et al. (2016), who observed an inverse relationship between the contribution of the FR to dermal permeability and $\log K_{o/w}$ for a set of compounds with varied molecular weights, binding properties and ionization characteristics. Frum et al (2007) recorded an initial increasing trend followed by a turning point and a sharp decrease. However, their experimental observations show a high $CFR$ for highly hydrophilic compounds, and near-zero values for compounds with high $\log K_{o/w}$. In light of their results, it is likely that the FR was filled with water, or with a high proportion of it, in both *in vitro* experiments reported, as opposed to expected conditions *in vivo*.

Figure 4 shows the results of the check on the uncertainty introduced by Mitragotri's expression. The graph shows that the trend does not change regardless of whether experimental measurements (Chen et al., 2013) or Mitragotri's prediction of the SC permeability is used to calculate the diffusion coefficient in the SC according to the methodology in Table 3. This simulated data was obtained for a set of actual chemicals with a range of values of molecular weight, protein binding properties and ionization characteristics for a water FR. The trend suggests thus that a water FR favours the permeation of hydrophilic compounds regardless of the other parameters, hence providing evidence that the trend observed in Figure 3 is not altered by the uncertainty introduced by using Mitragotri's QSPR to predict the permeability of the SC. This uncertainty check provides further evidence that the experimental *in vitro* results of Mohd et al. (2016) might have been obtained with a FR filled with a watery substance.

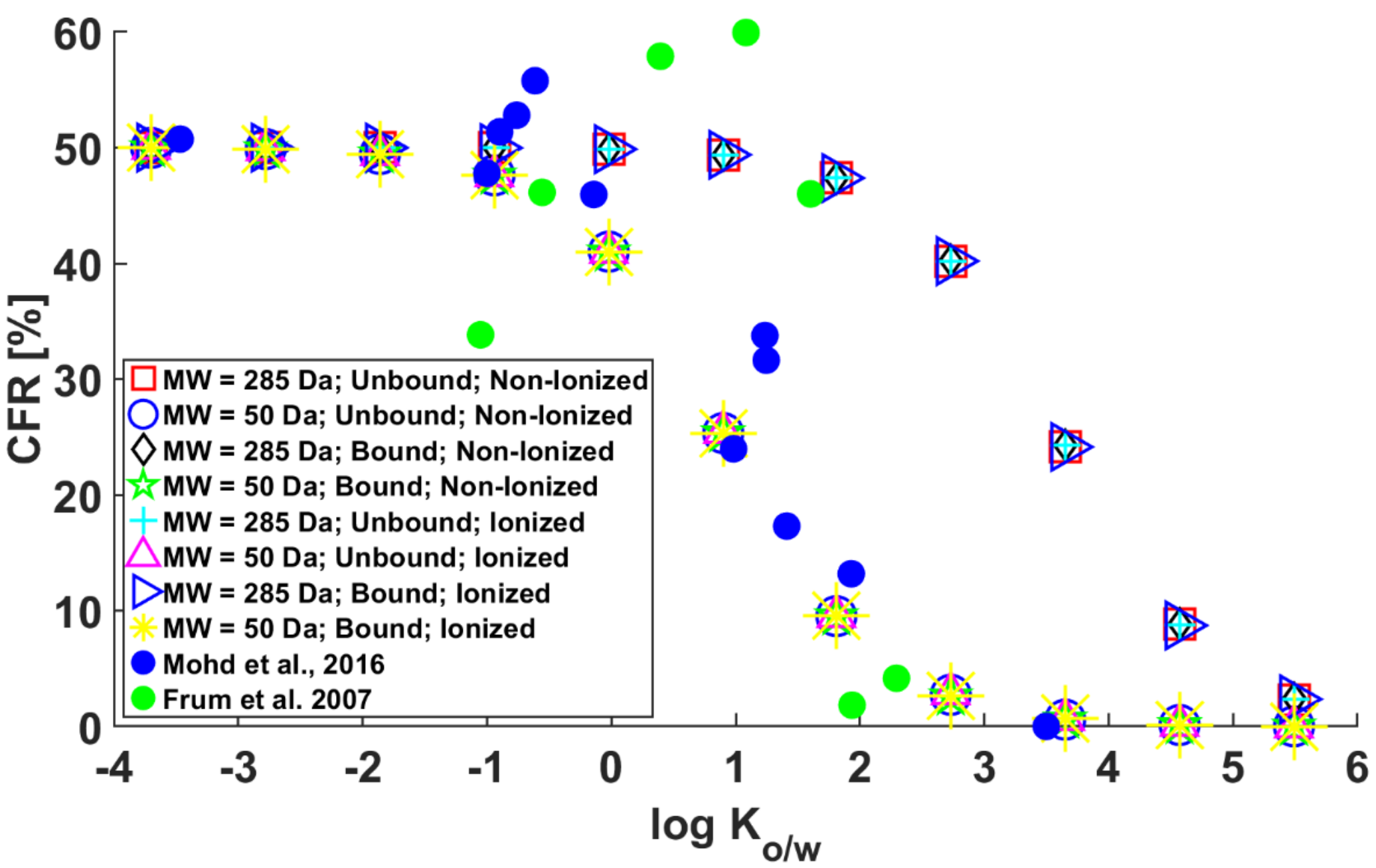

*Figure 3 Percentage of contribution of the follicular route CFR against log $K_{o/w}$ for water FR.*

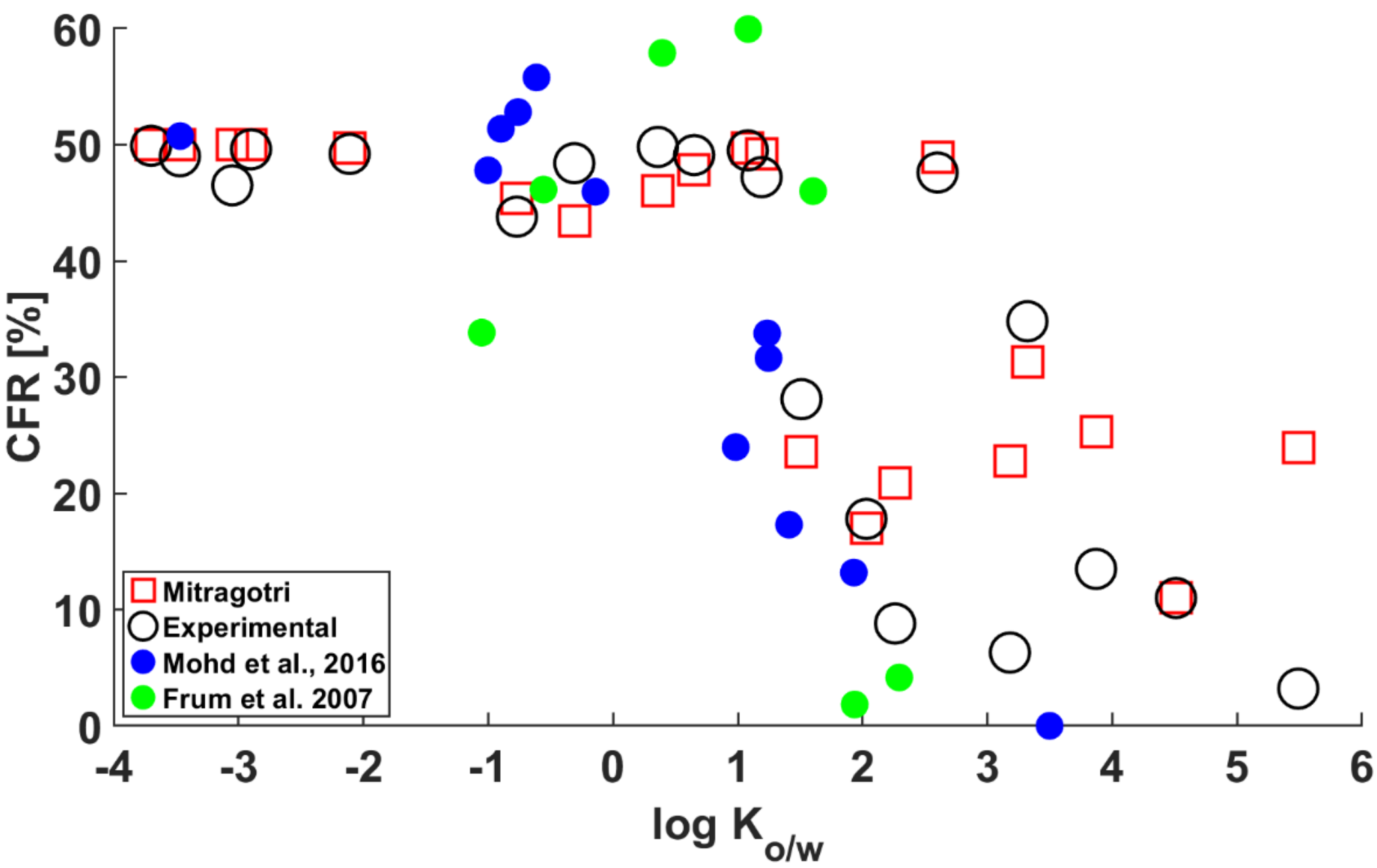

*Figure 4 Percentage of contribution of the follicular route CFR against log $K_{o/w}$ for Mitragotri's SC permeability uncertainty check.*

A further check was carried out to compare the simulated skin permeability with experimental measurements of the SC permeability for the set of chemical compounds in Table 4. The results of the comparison are included in Table 6. The simulated permeability of the skin with closed FR is the closest to the experimental data among the three simulated series, with slightly smaller values due to adding the contribution of the viable epidermis and dermis. The skin permeability increases notably for the results with impaired FR (i.e., flux restricted to half that of the open hair follicle) and open FR, hence suggesting the important role of the FR to overall dermal permeation.

*Table 6 Comparison between experimental measurement of the permeability of the SC and the simulated permeabilities of the skin for closed FR (no flux boundary between vehicle and FR), impaired FR (flux through concentration boundary of FR limited to half) and open FR.*

| Compound | Experimental $\log K_p$ [cm/s] [Chen et al., 2013] | Simulated Skin Permeability $\log K_{skin}$ [cm/s] for closed FR | Simulated Skin Permeability $\log K_{skin}$ [cm/s] for impaired FR | Simulated Skin Permeability $\log K_{skin}$ [cm/s] for open FR |
|---|---|---|---|---|
| Sucrose | -8.84 | -8.89 | -6.23 | -5.93 |
| Aspartic acid | -7.43 | -7.47 | -6.06 | -5.77 |
| Lysine | -6.87 | -6.92 | -6.04 | -5.76 |
| Histidine | -7.82 | -7.86 | -6.10 | -5.80 |
| Urea | -7.39 | -7.42 | -5.94 | -5.64 |
| Methanol | -6.38 | -6.41 | -5.75 | -5.50 |
| Ethanol | -7.06 | -7.09 | -5.88 | -5.60 |
| Nicotinic acid | -8.18 | -8.22 | -6.06 | -5.76 |
| Barbital | -7.51 | -7.56 | -6.12 | -5.82 |
| Aldosterone | -7.86 | -7.92 | -6.23 | -5.94 |
| Codeine | -7.09 | -7.15 | -6.17 | -5.89 |
| Pentanol | -5.78 | -5.84 | -5.63 | -5.49 |
| Hexanol | -5.44 | -5.54 | -5.43 | -5.35 |
| Salicylic acid | -5.07 | -5.27 | -5.22 | -5.18 |
| Hydrocortisone methylsuccinate | -7.23 | -7.29 | -6.25 | -5.97 |
| Naproxen | -4.97 | -5.25 | -5.22 | -5.19 |
| Testosterone | -6.21 | -6.28 | -5.95 | -5.76 |
| Progesterone | -5.43 | -5.59 | -5.51 | -5.45 |
| Diclofenac | -5.30 | -5.48 | -5.42 | -5.37 |
| Hydrocortisone octanoate | -4.76 | -5.21 | -5.19 | -5.18 |

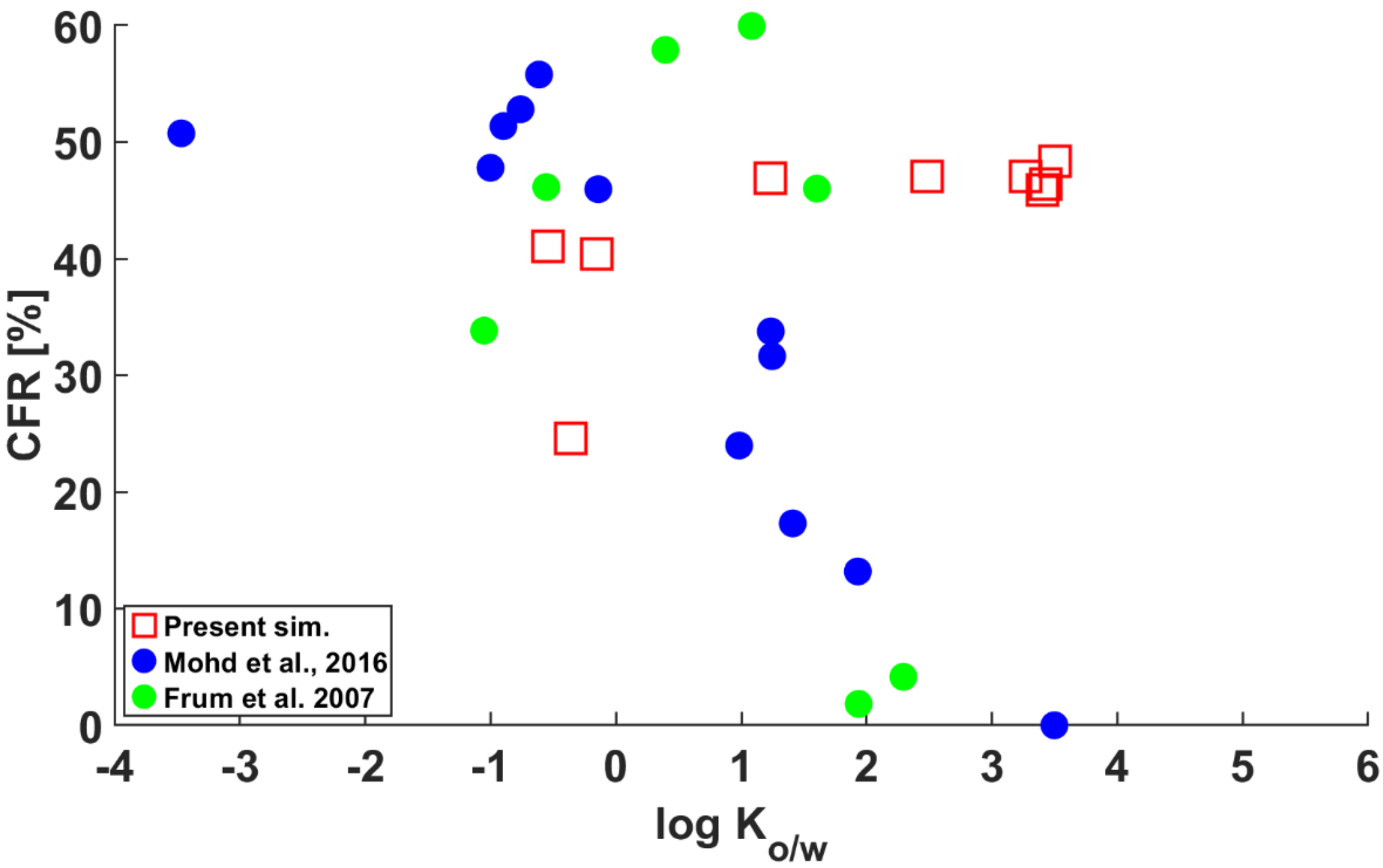

*Figure 5 Percentage of contribution of the follicular route CFR against $\log D_{o/w}$ for sebum FR.*

The results for a FR filled with sebum are included in Figure 5. The simulated results with a sebum FR do not follow the trend of the experimental data. Instead, permeants with low $\log K_{o/w}$ present a lower value of $CFR$ than those with the highest $\log K_{o/w}$. This is particularly clear for the case of theobromine ($\log K_{o/w}$ = -0.36), which presents a low $K_{se/w}$, hence reinforcing the hypothesis that the FR behaves as a hydrophilic shortcut *in vitro* and as a lipophilic shortcut *in vivo*. The simulations reveal that the FR plays an important role in overall permeation, with values of $CFR$ above 20% in all instances. Given the notable contribution of the FR in overall dermal permeation, these results highlight the importance of careful characterisation of the FR and the permeant before experimental work to avoid discrepancies between conclusions from results obtained *in vitro* and results obtained *in vivo*.

## Conclusions

An *in silico* study has been carried out to analyse the role that the follicular route has on overall dermal permeability, as information in the literature remains inconclusive regarding whether the follicular route favours the permeation of hydrophilic or lipophilic compounds. The *in silico* methodology in this study allows the analysis of a wide range of potential dermal permeation scenarios by varying parameters that describe the follicular route and the permeants.

The simulated data shows that the contribution of the follicular route can range from negligible to notable if compared to the alternative epidermal route depending mainly on the appropriate combination of lipophilicity/hydrophilicity of both the substance filling the follicular route and the compound being administered. Increased lipophilicity of the permeant results in greater contribution of the follicular route when filled with sebum as opposed to cases with a follicular route filled with water. The molecular weight of the permeant proves important, too, with the follicular route gaining relevance for permeants with increased molecular weight. The simulated trends obtained between $\log K_{o/w}$ and the contribution of the follicular route agree with experimental work *in vitro*, suggesting that the experimental data were obtained with a follicular route filled with a watery substance. Conversely, dermal products are designed to be administered with a follicular route filled with sebum. The effect of switching to a sebum follicular route in the simulation has a large impact on the results, hence opening the way towards distorted conclusions from *in vitro* experiments.

As a result, uncontrolled modifications of the physico-chemical properties of the follicular route between *in vivo* and *in vitro* experimental conditions due to sample manipulation and cessation of vital functions may give rise to distorted interpretation of dermal permeation tests. Thus, determination of the composition of the substance that fills the follicular route would be advisable prior to dermal permeation experimental work.

**Data availability:** Data available in the following link: OSF | Follicular route dermal permeation Simulation